\documentclass{PoS_arXiv}

\usepackage{amsmath}
\usepackage{graphicx}
\usepackage{url}
\usepackage{epsfig}
\usepackage{cite}
\usepackage{wrapfig}

\newcommand{\sq}{{\tilde{q}}}
\newcommand{\sqb}{{\bar{\tilde{q}}}}
\newcommand{\gl}{{\tilde{g}}}
\renewcommand{\d}{\mathrm{d}}

\title{NNLL resummation for squark-antisquark production}

\ShortTitle{NNLL resummation for squark-antisquark production}

\author{Wim Beenakker$^a$, Silja Brensing$^b$, Michael Kr\"amer$^c$, Anna Kulesza$^c$, Eric Laenen$^d$ and \speaker{Irene Niessen}$^a$\\
\llap{$^a$} IMAPP, Radboud University Nijmegen, P.O. Box 9010 NL-6500 GL Nijmegen, The Netherlands.\\
\llap{$^b$} DESY, Theory Group, Notkestrasse 85, D-22603 Hamburg, Germany.\\
\llap{$^c$} Institute for Theoretical Particle Physics and Cosmology, RWTH Aachen University D-52056 Aachen, Germany.\\
\llap{$^d$} ITFA, University of Amsterdam, Science Park 904, 1018 XE, Amsterdam, \\
ITF, Utrecht University, Leuvenlaan 4, 3584 CE Utrecht\\
Nikhef Theory Group, Science Park 105, 1098 XG Amsterdam, The Netherlands\\
E-mail:  \email{w.beenakker@science.ru.nl}, \\
\email{silja.christine.brensing@desy.de}, \\
\email{mkraemer@physik.rwth-aachen.de}, \\
\email{anna.kulesza@physik.rwth-aachen.de}, \\
\email{t45@nikhef.nl},\\
\email{i.niessen@science.ru.nl}}

\abstract{We report on the resummation of soft-gluon emissions for squark-antisquark production at next-to-next-to-leading-logarithmic (NNLL) accuracy. We will put particular emphasis on the one loop hard matching coefficients required to perform the resummation. Furthermore we will discuss the numerical effect of the different ingredients in the corrections. We find a significant reduction in the scale uncertainty and a considerable increase in the prediction of the total cross section at the central scale. Compared to the next-to-leading order prediction, the corrections increase the cross section by up to 30\% for 1.5~TeV squarks at a centre-of-mass (CM) energy of 7~TeV at the LHC.}

\FullConference{10th International Symposium on Radiative Corrections (Applications of Quantum Field Theory to Phenomenology)\\
                September 26-30, 2011\\
                Mamallapuram, India}

\begin{document}

\section{Introduction}
\label{s:intro}

Supersymmetry (SUSY), and in particular the minimal supersymmetric Standard Model\cite{Nilles:1983ge}, can provide a solution to the hierarchy problem, accommodate gauge coupling unification and offer a dark matter candidate. Since these issues are addressed only if the SUSY scale is comparable to the weak scale, one would expect new SUSY particles (sparticles) with masses in the TeV range, which could be measured at the Large Hadron Collider (LHC). Particularly the coloured sparticles, squarks ($\tilde q$) and gluinos ($\tilde g$), would be pair-produced copiously in hadronic collisions and thus offer the strongest sensitivity. Searches at the LHC with a centre-of-mass (CM) energy of $\sqrt{S}=7$~TeV have placed lower limits on squark and gluino masses around 1~TeV \cite{Aad:2011ib,Chatrchyan:2011zy}.

The leading order (LO) predictions for inclusive squark and gluino hadroproduction depend strongly on the renormalization and factorization scale. This dependence is reduced significantly if higher-order SUSY-QCD corrections are included. If the renormalization and factorization scales are chosen close to the average mass of the produced particles, the corrections generally increase the size of the cross section compared to the LO prediction. Consequently, the SUSY-QCD corrections have a substantial impact on the determination of mass exclusion limits and lead to a significant reduction of the uncertainties on SUSY mass or parameter values in the case of discovery \cite{Baer:2007ya,Dreiner:2010gv}. 

The squark-antisquark production processes have been known for quite some time at next-to-leading order (NLO) in SUSY-QCD \cite{Beenakker:1994an,Beenakker:1996ch}.
A significant part of the NLO QCD corrections can be attributed to the threshold region, where the partonic CM energy is close to the kinematic production threshold and the NLO corrections are dominated by soft-gluon emission off the coloured particles in the initial and final state. These soft-gluon corrections can be taken into account to all orders in perturbation theory by means of threshold resummation techniques \cite{Sterman:1986aj,Catani:1989ne}.
This has been done for all MSSM squark and gluino production processes at next-to-leading-logarithmic (NLL) accuracy \cite{Kulesza:2008jb,Kulesza:2009kq,
Beenakker:2009ha,Beenakker:2010nq,Beenakker:2011fu}. 

Recently, resummation at NNLL accuracy for squark-anti\-squark pair production was presented~\cite{Beenakker:2011sf}. Compared to the NLL calculation the new ingredients are the one-loop matching coefficients, which contain the NLO cross section near threshold, and the two-loop soft anomalous dimensions. 
We will discuss the impact of the corrections and provide an estimate of the theoretical uncertainty due to scale variation.

In section~\ref{s:resummation} we discuss NNLL resummation for squark-antisquark pair-production. In section~\ref{s:Ccoeff} we present the calculation of the hard matching coefficients. The numerical results for the LHC with a CM energy of $\sqrt{S}=7$~TeV are presented in section~\ref{s:numres}. We conclude in section~\ref{s:conclusion}.

\section{Threshold resummation at NNLL}
\label{s:resummation}

\begin{figure}[!h]
\begin{center}
\includegraphics[width=0.65\textwidth]{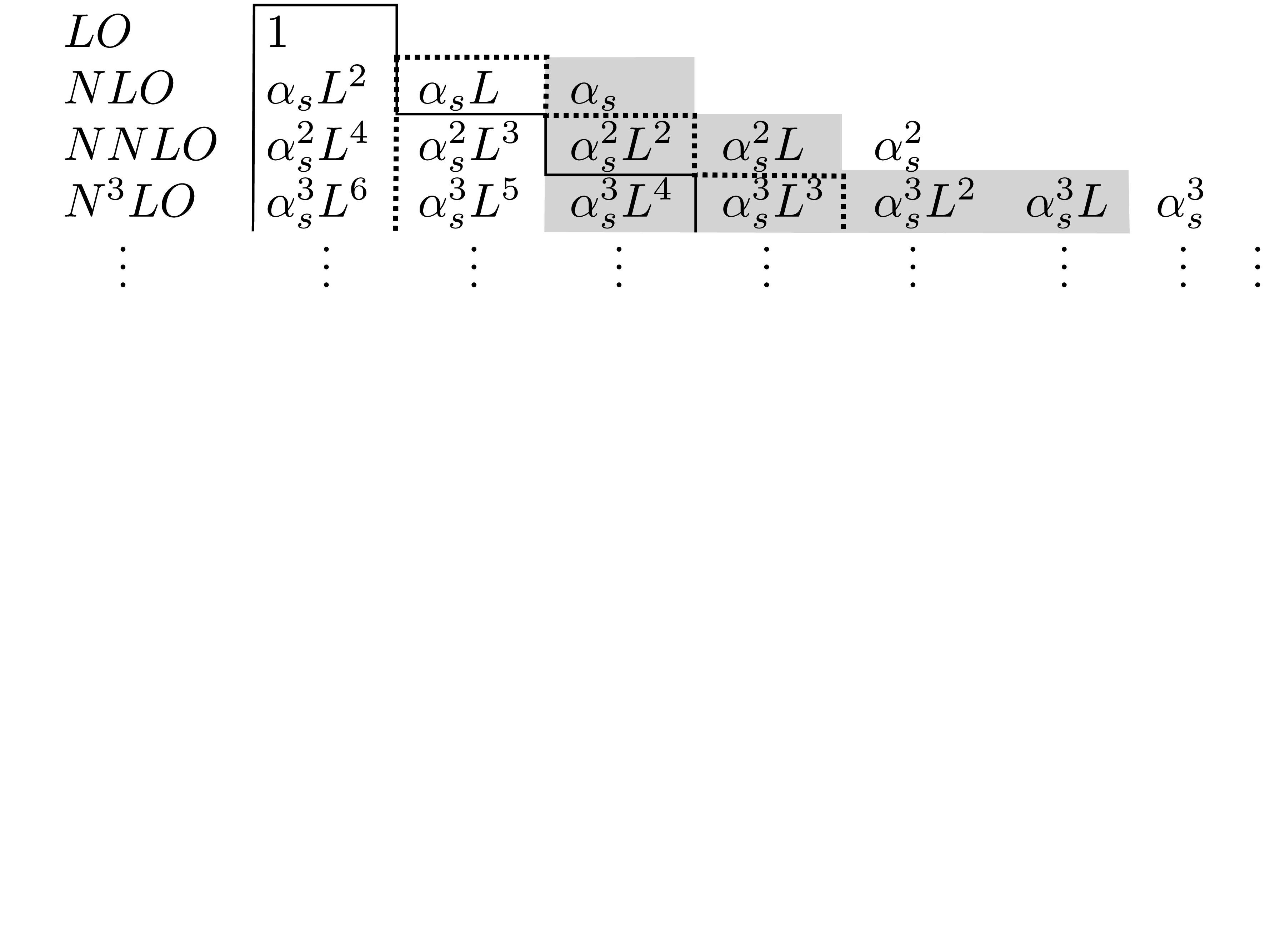}
\end{center}
\caption{Schematic form of the logarithmically enhanced terms with the LO cross section factored out. The terms enclosed in the solid line occur in the LL approximation, which completes the first column. The new terms in the NLL approximation are enclosed in the dotted line and complete the second column. The shaded region corresponds to new terms coming from the NNLL approximation, which completes the third and fourth column. We resum the logarithms to all orders, so the columns extend downwards to infinity.\label{fig:expansion}}
\end{figure}

We first briefly review the formalism of threshold resummation for $\sq\sqb$ production. The inclusive cross section $\sigma_{h_1h_2\rightarrow \sq\sqb}$ can be written in terms of its partonic version
$\sigma_{ij\rightarrow \sq\sqb}$ as:
\[  \sigma_{h_1 h_2 \to \sq\sqb}\bigl(\rho, \{m^2\}\bigr) 
  =\sum_{i,j} \int\! d x_1 d x_2\,d\hat{\rho}\,
        \delta\!\!\left(\!\hat{\rho} - \frac{\rho}{x_1 x_2}\right)
 f_{i/h_{1}}(x_1,\mu^2 )\,f_{j/h_{2}}(x_2,\mu^2 )
        \sigma_{ij \to \sq\sqb}\bigl(\hat{\rho},\{ m^2\},\mu^2\bigr)\,,\]
where $\{m^2\}$ denotes all masses entering the calculations, $i,j$ are the initial parton flavours, $f_{i/h_1}$ and $f_{j/h_2}$ the parton distribution functions, $\mu$ is the common factorization and renormalization scale and $x_{1,2}$ are the momentum fractions of the partons in hadrons $h_1$ and $h_2$. The hadronic threshold for the inclusive production of two squarks with mass $m_\sq$ corresponds to a hadronic CM energy squared $S=4m_\sq^2$. Therefore we define a threshold variable $\rho=4m_\sq^2/S$ and its partonic equivalent $\hat{\rho}=4m_\sq/s$, with $s=x_1x_2S$ the partonic CM energy squared. The resummation of the soft-gluon contributions is performed after taking a Mellin transform (indicated by a tilde) of the cross section,
\begin{align}
  \label{eq:10}
  \tilde\sigma_{h_1 h_2 \to \sq\sqb}\bigl(N, \{m^2\}\bigr) 
  &\equiv \int_0^1 d\rho\;\rho^{N-1}\;
           \sigma_{h_1 h_2\to \sq\sqb}\bigl(\rho,\{ m^2\}\bigr) \nonumber\\
  &=      \;\sum_{i,j} \,\tilde f_{i/{h_1}} (N+1,\mu^2)\,
           \tilde f_{j/{h_2}} (N+1, \mu^2) \,
           \tilde{\sigma}_{ij \to \sq\sqb}\bigl(N,\{m^2\},\mu^2\bigr)\,.
\end{align}
The threshold limit $\hat\rho\to1$ corresponds to $N\to\infty$. Near threshold, fixed-order perturbation theory does not converge well due to logarithmically enhanced terms. By resumming large logarithmic corrections containing $L=\log(N)$ to all orders, a new perturbative expansion arises, where first the leading logarithmic (LL) corrections are resummed, followed by the NLL, NNLL, ... contributions. The relation between this new perturbative expansion and fixed-order perturbation theory is shown schematically in Figure~\ref{fig:expansion}.

The all-order summation of these logarithmic terms is based on the
near-threshold factorization of different classes of radiation: hard,
(soft)-collinear, and wide-angle soft radiation
\cite{Sterman:1986aj,Catani:1989ne,Bonciani:1998vc,Contopanagos:1996nh,
  Kidonakis:1998bk,Kidonakis:1998nf}. Near threshold the resummed partonic cross section to NNLL accuracy has the form \cite{Sterman:1986aj,Catani:1989ne,Beneke:2010da}:
\begin{align}
  \label{eq:12}
  \tilde{\sigma}^{\rm (res)} _{ij\rightarrow \sq\sqb}\bigl(N,\{m^2\},&\mu^2\bigr) 
  =\sum_{I}\,
      \tilde\sigma^{(0)}_{ij\rightarrow \sq\sqb,I}\bigl(N,\{m^2\},\mu^2\bigr)\, \exp\Big[L g_1(\alpha_{\rm s}L) + g_{2,I}(\alpha_{\rm s}L) + \alpha_{\rm s}g_{3,I}(\alpha_{\rm s}L) \Big] \nonumber\\
  & \times\,\left(1+\frac{\alpha_{\rm s}}{\pi}\;{\cal C}^{\rm Coul,(1)}_{ij\to\sq\sqb,I}(N,\{m^2\},\mu^2)\right)\;\left(1+\frac{\alpha_{\rm s}}{\pi}\;{\cal C}^{\rm (1)}_{ij\to\sq\sqb,I}(\{m^2\},\mu^2)\right)\,,
\end{align}
where $\tilde{\sigma}^{(0)}_{ij \rightarrow \sq\sqb,I}$ are the colour-decomposed LO cross sections in Mellin-moment space. The colour label $I$ corresponds to an irreducible representation of the colour structure of the process, which for squark-antisquark production can be either a singlet or an octet~\cite{Beneke:2009rj,Kulesza:2008jb,Kulesza:2009kq}. The exponent in the first line of Eq.~\eqref{eq:12} captures all dependence on the large logarithm $L$. The last line contains the one-loop Coulomb contribution ${\cal C}^{\rm Coul,(1)}$ and hard matching coefficient~${\cal C}^{(1)}$. 

Setting the hard matching and Coulomb coefficients to 0 and keeping only the $g_1$ term in Eq.~\eqref{eq:12} constitutes the LL approximation. Including the $g_2$ term as well corresponds to NLL. For NNLL accuracy also the $g_3$ term and the one-loop hard matching and Coulomb coefficients need to be taken into account, as can be seen explicitly by expanding Eq.~\eqref{eq:12} and comparing it to figure~\ref{fig:expansion}. The $g_3$ term can be found in~\cite{Moch:2005ba,Czakon:2009zw,Moch:2008qy,Beenakker:2011sf} and the ingredients for ${\cal C}^{\rm Coul,(1)}_{ij\to\sq\sqb,I}$ in~\cite{Catani:1996dj,Fadin:1990wx,Beenakker:2011sf}. The calculation of the NLO hard matching coefficients ${\cal C}^{\rm (1)}_{ij\to\sq\sqb,I}$ will be discussed in section~\ref{s:Ccoeff}.

The hadronic cross section $\sigma_{h_1 h_2 \to \sq\sqb}$ can be recovered by taking the inverse Mellin transform. We retain the information contained in the complete NLO cross section~\cite{Beenakker:1996ch} by combining the NLO and NNLL results through a matching procedure that avoids double counting of the NLO terms:
\begin{align}
 \label{eq:matching}
  &\sigma^{\rm (NLO+NNLL~matched)}_{h_1 h_2 \to \sq\sqb}\bigl(\rho, \{m^2\},\mu^2\bigr) 
  = \sum_{i,j}\,\int_\mathrm{CT}\,\frac{dN}{2\pi i}\,\rho^{-N} \tilde f_{i/h_1}(N+1,\mu^2)\tilde f_{j/h_{2}}(N+1,\mu^2) \\
&\qquad\times       \left[\tilde\sigma^{\rm(res,NNLL)}_{ij\to \sq\sqb}\bigl(N,\{m^2\},\mu^2\bigr)
             \,-\, \tilde\sigma^{\rm(res,NNLL)}_{ij\to \sq\sqb}\bigl(N,\{m^2\},\mu^2\bigr)
       {\left.\right|}_{\scriptscriptstyle{\rm (NLO)}}\, \right]  +\sigma^{\rm (NLO)}_{h_1 h_2 \to \sq\sqb}\bigl(\rho, \{m^2\},\mu^2\bigr).  \nonumber
\end{align}
We adopt the ``minimal prescription'' of Ref.~\cite{Catani:1996yz} for
the contour CT of the inverse Mellin transform in Eq.~(\ref{eq:matching}).
In order to use standard parametrizations of parton distribution
functions in $x$-space we employ the method introduced in
Ref.~\cite{Kulesza:2003wn}.

\section{Calculation of the hard matching coefficients}
\label{s:Ccoeff}

We now discuss the calculation of the one-loop hard matching coefficients ${\cal C}^{(1)}$. The calculations were done using \texttt{FORM} \cite{Vermaseren:2000nd}. After performing an expansion of the
NLO cross section in $\beta$, the hard matching coefficients ${\cal
  C}^{\rm(1)}$  are determined by the terms in the NLO cross section
that are proportional to $\beta$,  $\beta \log(\beta)$ and  $\beta \log^2(\beta)$. Terms that contain higher powers of $\beta$ are suppressed by powers of $1/N$ in Mellin-moment space and do not contribute to the matching coefficient. In contrast to the case of top-pair production in Ref.~\cite{Czakon:2008cx}, there is no full analytic result for the real corrections to squark-antisquark production, so we cannot take the explicit threshold limit.

The virtual corrections for squark-antisquark production can be obtained from the full analytic calculation as presented in Ref.~\cite{Beenakker:1996ch}. First we need to colour-decompose the result. We only need the colour decomposition of the LO matrix element. Due to the orthogonality of the $s$-channel colour basis that we use, the full matrix element squared is then automatically colour-decomposed:
\[|{\cal M}|^2_{{\rm NLO},I}=2\mathrm{Re}({\cal M}_{\rm NLO}{\cal M}^*_{{\rm LO},I}).\]
Then we are left with an expression in terms of masses, Mandelstam variables and scalar integrals. Since we need the cross section to ${\cal O}(\beta)$, we have to expand $|{\cal M}|^2$ to zeroth order in $\beta$ to obtain the virtual part of the hard matching coefficients.

The integrated real corrections at threshold are formally phase-space suppressed near threshold unless the integrand compensates this suppression. Therefore we can construct the real corrections at threshold from the singular behaviour of the matrix element squared, which can be obtained using dipole subtraction \cite{Catani:1996vz,Catani:2002hc}. First we recall that the cross section can be split into three parts: a part with three-particle kinematics $\sigma^{\{3\}}$, one with two-particle kinematics $\sigma^{\{2\}}$, and a collinear counterterm $\sigma^C$ to remove the initial-state collinear singularities. These parts are well-defined in $n=4-2\epsilon$ dimensions, but their constituents diverge for $\epsilon\to0$. With the aid of an auxiliary cross section $\sigma^A$, which captures all singular behaviour, all parts are made finite and integrable in four space-time dimensions. This auxiliary cross section is subtracted from the real corrections $\sigma^{\rm R}$ at the integrand level to obtain $\sigma^{\{3\}}$ and added to the virtual corrections $\sigma^{\rm V}$, which defines $\sigma^{\{2\}}$:
\[\sigma^{\rm NLO}=\int_{3}\big[\d\sigma^{\rm R}-\d\sigma^{\rm A}\big]_{\epsilon=0}+\int_{2}\big[\d\sigma^{\rm V}+\int_1\d\sigma^{\rm A}\big]_{\epsilon=0}+\sigma^{\rm C}\equiv\sigma^{\{3\}}+\sigma^{\{2\}}+\sigma^{\rm C}\]
Compared to the case of two-parton kinematics, the phase space of $\sigma^{\{3\}}$ is limited by the energy of the third, radiated massless particle. Near the two-particle threshold, the maximum energy of the radiated particle, and thus the available phase space, equals $E_{\rm max}=\sqrt{s}-2m_\sq\propto\beta^2$. Since after subtracting $\sigma^A$ no divergences are left in the integrand of $\sigma^{\{3\}}$, the leading contribution of $\sigma^{\{3\}}$ is at most proportional to $\beta^2$ and can thus be neglected. So at threshold the real radiation is completely specified by the singular behaviour contained in $\sigma^A$ and can therefore be determined by summing over dipoles that correspond to pairs of ordered partons \cite{Catani:2002hc} and taking the threshold limit.

\begin{figure}[!h]
\hspace{-0.55cm}
\begin{tabular}{lll}
(a)\hspace{-0.3cm}\epsfig{file=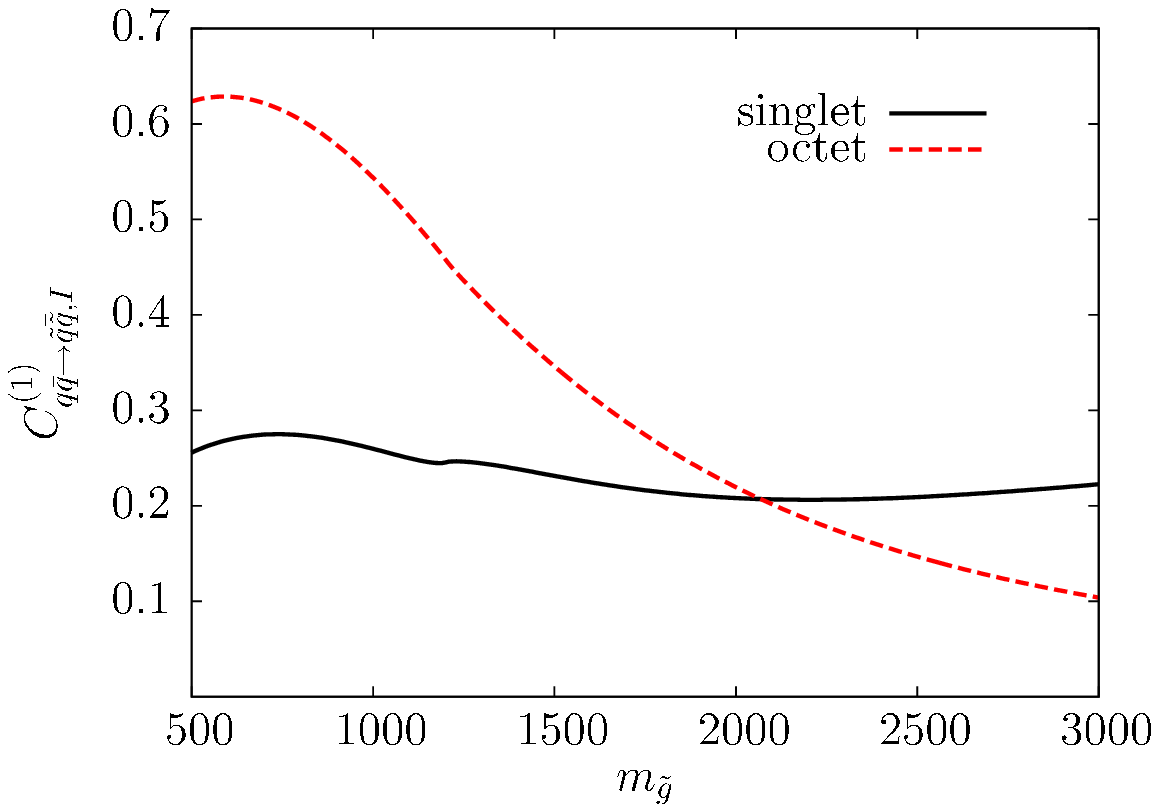, width=0.47\columnwidth}&
(b)\hspace{-0.3cm}\epsfig{file=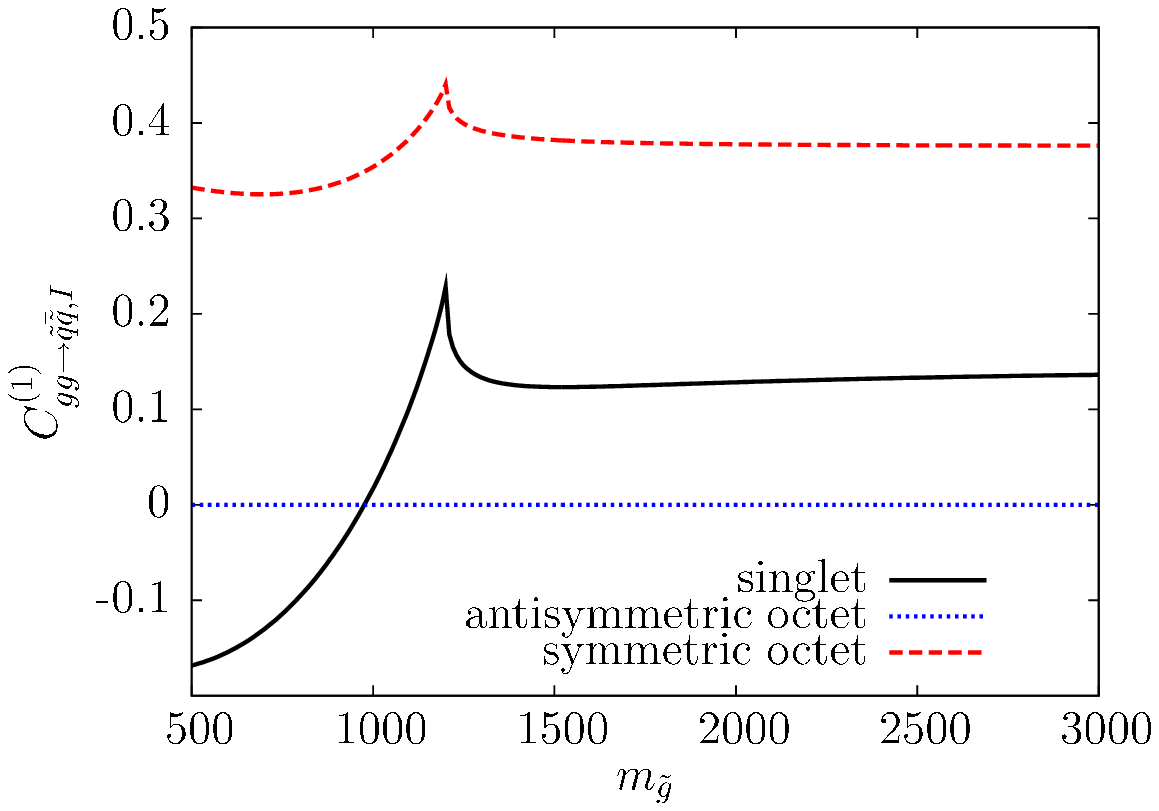, width=0.47\columnwidth}
\end{tabular}
\caption{Gluino-mass dependence of the colour-decomposed NLO hard matching coefficients for the $q\bar q$ initiated channel (a) and the $gg$ initiated channel (b). We have set $\mu=m_\sq=1.2$~TeV and $m_t=172.9$~GeV.\label{fig:Ccoeffplots}}
\end{figure}

After combining the real and the virtual corrections the hard matching coefficients can be obtained by taking the Mellin transform and omitting the Coulomb corrections and the $\log(N)$ terms. The complete expressions for the hard matching coefficients of the squark-antisquark production processes can be found in Ref.~\cite{Beenakker:2011sf}. Their behaviour for varying gluino mass is shown in Fig.~\ref{fig:Ccoeffplots}.

\section{Numerical results}
\label{s:numres}

We now present numerical results for NNLL-resummed squark-antisquark pair-production at the LHC for a CM energy of 7~TeV. We use the 2008 NLO MSTW parton distribution functions~\cite{Martin:2009iq} with the corresponding $\alpha_{\rm s}(M_{Z}^2) = 0.120$. We have used a top quark mass of $m_t=172.9$~GeV~\cite{PDG}. In order to study the effects from the hard matching coefficients and the Coulomb corrections separately, we compare the NLL matched cross section $\sigma^{\rm NLO+NLL}$, which is based on the calculations presented in \cite{Beenakker:2009ha,Kulesza:2008jb,Kulesza:2009kq}, the NNLL matched cross section without the Coulomb contributions to the resummation $\sigma^{\rm NLO+NNLL\; w/o\; Coulomb}$, which has ${\cal C}^{\rm Coul,(1)}=0$ in Eq.~\eqref{eq:12}, and the NNLL matched cross section $\sigma^{\rm NLO+NNLL}$, which does include the Coulomb correction ${\cal C}^{\rm Coul,(1)}$. All cross sections are matched to the NLO result calculated with {\tt PROSPINO}~\cite{Beenakker:1996ch,prospino} using Eq.~\eqref{eq:matching}.

\begin{figure}[!h]
(a)\includegraphics[width=0.45\textwidth]{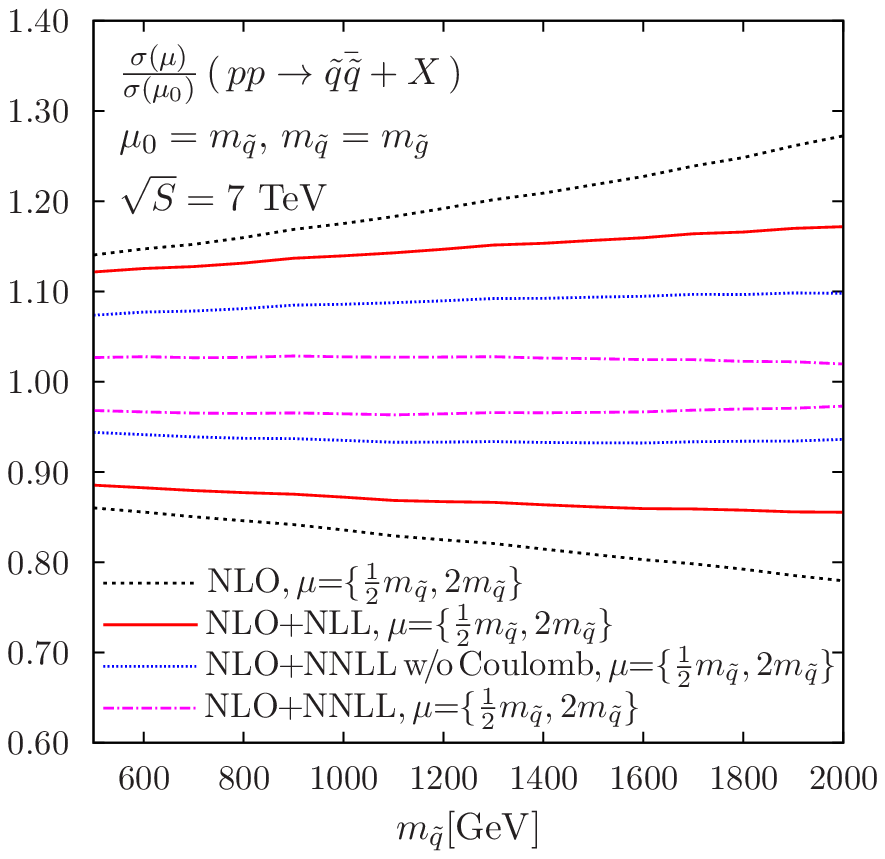}\hfill
(b)\includegraphics[width=0.45\textwidth]{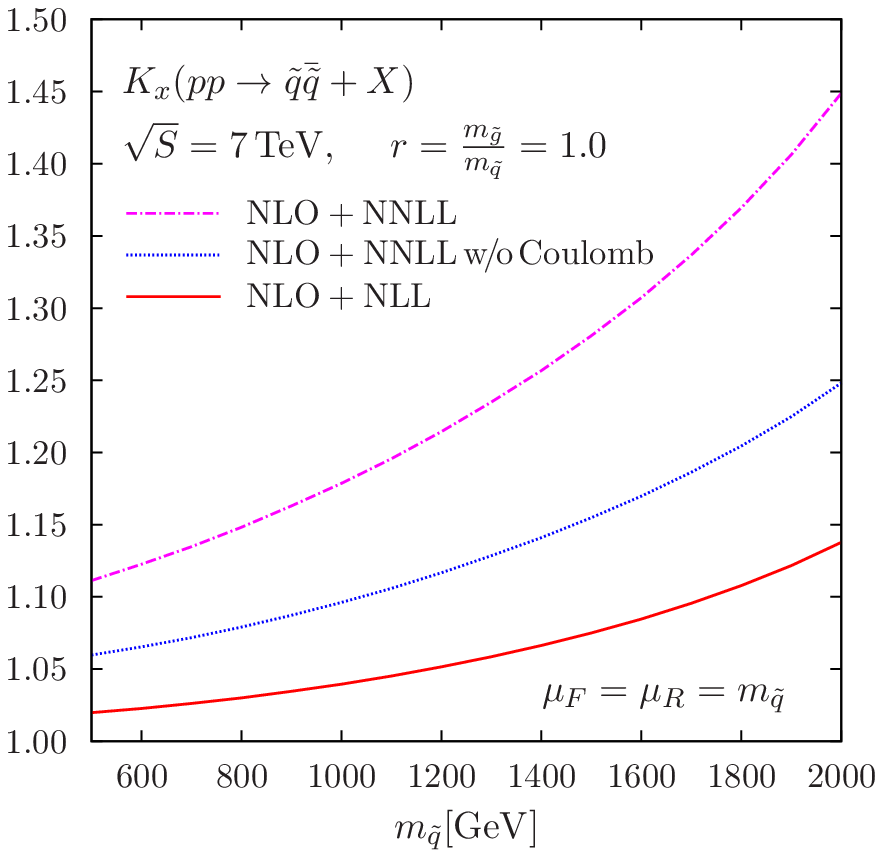}
\caption{The scale uncertainty (a) and K-factor (b) of the LO, NLO, NLO+NLL and NLO+NNLL (both with the Coulomb part ${\cal C}^{\rm Coul,(1)}$ and without it) squark-antisquark cross sections for the LHC at 7 TeV. The central scale and the gluino mass have been taken equal to the squark mass. In (a) the common renormalization and factorization scale has been varied in the range $m_\sq/2\le\mu\le 2m_\sq$.\label{fig:scale}}
\end{figure}

Figure \ref{fig:scale}(a) shows the mass dependence of the scale uncertainty for the squark-antisquark cross section. The squark and gluino mass have been taken equal and the scale has been varied in the range $m_\sq/2\le\mu\le 2m_\sq$.
We see that the scale uncertainty reduces when higher-order corrections are included. In this range of squark masses, the NNLL resummation without the Coulomb corrections ${\cal C}^{\rm Coul,(1)}$ already reduces the scale uncertainty to at most 10\%. The inclusion of the Coulomb term ${\cal C}^{\rm Coul,(1)}$ in the resummed NNLL prediction results in a scale uncertainty of only a few percent.

In figure~\ref{fig:scale}(b) we study the mass dependence of the $K$-factor $K_x=\sigma^x/\sigma^{\rm NLO}$ at the central scale $\mu=m_\sq$ and $m_\gl=m_\sq$. The 
theoretical prediction of the cross section increases as more
corrections are included. The effect is more pronounced for
higher masses, since in that case particles are produced closer to threshold. The NNLL resummation without the Coulomb
corrections ${\cal C}^{\rm Coul,(1)}$ already results in a 25\%
increase of the cross section with respect to the NLO cross section
for squarks of 2~TeV. The contribution from
the Coulomb term to the resummed NLO+NNLL cross section is even larger, yielding a total
$K$-factor of 1.45. Although the effect from the Coulomb
corrections could be somewhat smaller in reality due to the finite
lifetime of the squarks, figure~\ref{fig:scale}(b) suggests that the NNLL
contribution will remain large.

\section{Conclusions}
\label{s:conclusion}

We have discussed the NNLL resummation of threshold corrections for squark-antisquark hadroproduction and presented numerical results for the NLO+NNLL resummed cross section for squark-antisquark production at the LHC with a CM energy of 7 TeV. For a squark mass of 2 TeV, the
NLO+NNLL squark-antisquark cross section is up to 45\% larger than the corresponding
NLO cross section. The correction is reduced to 25\% if
the resummation contributions due to Coulombic interactions are omitted.
The scale dependence is reduced significantly, particularly after inclusion of the Coulomb corrections. However, the observed reduction in the scale dependence
might be modified somewhat by the inclusion of the width of the
particles or by matching to the full NNLO result, which is not
available. A very conservative estimate of the scale uncertainty is provided by the NLO+NNLL w/o Coulomb results, which do not include the Coulomb corrections.
This information could be used to improve current limits on SUSY masses or, in the case that SUSY is found, to more accurately determine the masses of the sparticles.

\section*{Acknowledgments}

This work has been supported in part by the Helmholtz Alliance
``Physics at the Terascale'', the Foundation for Fundamental
Research of Matter (FOM), program 104 ``Theoretical Particle Physics in
the Era of the LHC", the DFG SFB/TR9 ``Computational Particle Physics'',
and the European Community's Marie-Curie Research Training Network
under contract MRTN-CT-2006-035505 ``Tools and Precision Calculations
for Physics Discoveries at Colliders''.

\end{document}